\documentclass[preprint,5p,times,twocolumn]{elsarticle}

\usepackage{graphicx}

\usepackage{amssymb}

\usepackage{lineno}

\usepackage{upgreek}
\usepackage{color}

\journal{Nuclear Physics A}

\begin{document}

\begin{frontmatter}

\title{Gas cell density characterization for laser wakefield acceleration}

\author[label1]{T. L. Audet}
\ead{thomas.audet@u-psud.fr}
\author[label1]{P. Lee}
\author[label1]{G. Maynard}
\author[label2]{S. Dobosz Dufr\'enoy}
\author[label2]{A. Maitrallain}
\author[label2]{M. Bougeard}
\author[label2]{P. Monot}
\author[label1]{B. Cros}
\ead{brigitte.cros@u-psud.fr}
\address[label1]{LPGP, CNRS, Univ. Paris-Sud, Universit\'e Paris-Saclay, 91405, Orsay, France}
\address[label2]{LIDYL, CEA, Universit\'e Paris-Saclay, 91191, Gif-sur-Yvette, France.}

\begin{abstract}
In the design of laser plasma electron injectors for multi-stage laser driven wakefield accelerators, the control of plasma density is a key element to stabilize the acceleration process. A cell with variable parameters is used to confine the gas and tailor the density profile. The gas filling process was characterized both experimentally and by fluid simulations. Results show a good agreement between experiments and simulations. Simulations were also used to study the effect of each of the gas cell parameters on the density distribution and show the possibility to finely control the density profile.

\end{abstract}


\end{frontmatter}


\section{Introduction}
\label{S:1}

Laser wakefield accelleration (LWFA) relies on the interaction of a high intensity laser with a controlled under-dense plasma. The plasma can be generated by discharge \cite{Spence2000} or by laser field ionization of a gas medium \cite{Ammosov1986}. The medium can be either a gas jet \cite{Semushin2001}, or gas confined inside a capillary \cite{Andreev2010,Desforges2014} or a gas cell \cite{Clayton2010,Audet2016}. A laser guiding device, such as dielectric capillaries or plasma channels is usually needed to accelerate electrons to the GeV level in a single stage because of laser diffraction limiting the acceleration length. 
In the frame of the EuPRAXIA design study \cite{Walker2017} and the CILEX-Apollon project \cite{Cros2013}, multi-stage schemes are investigated. In these schemes, an injector generates very high quality electrons with energy of the order of few hundreds of MeV, before injection and acceleration in a subsequent plasma stage. For an injector of the order of 200 MeV, the acceleration length can be reduced to the mm scale for which a gas cell is a suitable confinement structure \cite{Lee2017}. For example, for injectors operating in a regime where injection of electrons is assisted by ionization of heavy atoms in a light gas background, the volume of injection can be controlled to improve the beam quality by using specific density distribution.
In this context, ELISA (ELectron Injector for compact Staged high energy Accelerator) was designed as a variable parameter gas cell to confine the gas target, and shape the plasma density profile to generate electron bunches suitable for multi-stage LWFA.
The remaining of the paper is organized as follows : Section 2 presents the gas cell design, Section 3 shows its experimental characterization using interferometric measurements and Section 4 describes the resulting density distribution for different simulation parameters.

\section{Cell description}
\label{S:2}
\begin{figure}[h]
\centering
\includegraphics[scale=0.55]{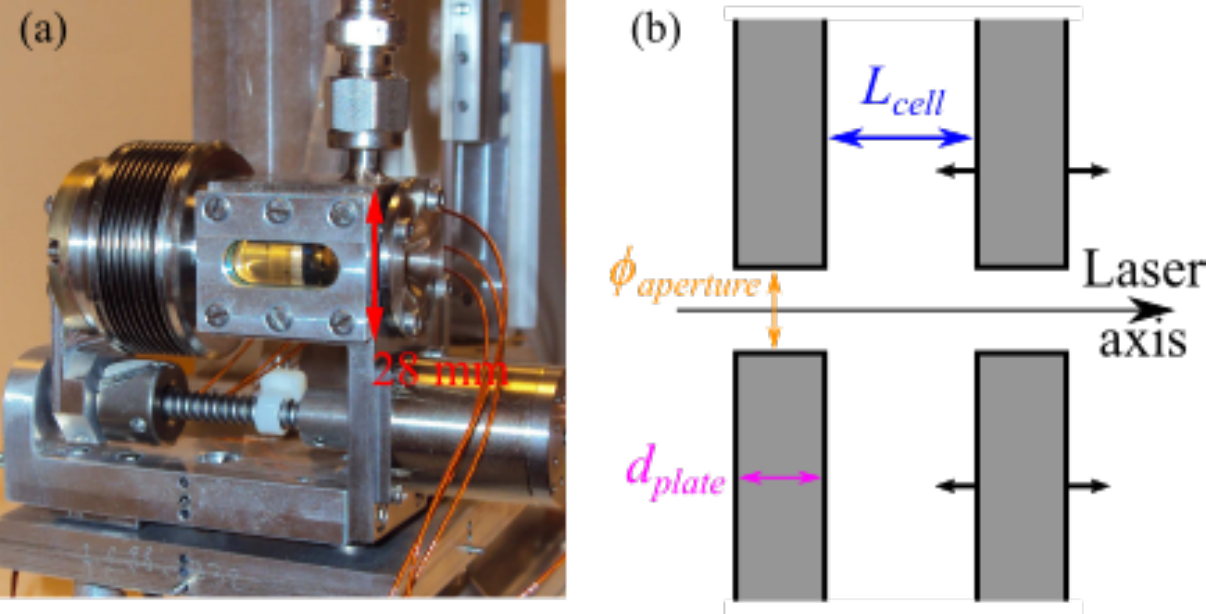}
\caption{(a) Photograph of the ELISA gas cell.(b) Schematic longitudinal cut of the gas cell. \label{fig:Fig1}}
\end{figure}

The ELISA gas cell used to confine the hydrogen gas ($\mathrm{H_2}$) is shown in Fig. \ref{fig:Fig1}(a). It consists in a 20 mm diameter cylinder with a 3 mm diameter gas inlet located on the side, close to the cell entrance. Metallic plates are placed at each end of the cylinder and apertures are drilled on these plates to allow the laser and electrons to go through during laser wakefield experiments.

 In the following the thickness of the plates will be denoted by $d_{plate}$ and the diameter of the apertures will be denoted $\phi_{aperture}$ as illustrated in Fig. \ref{fig:Fig1}(b). The plates are replaceable and are typically replaced every day before experiments with a high-power laser. 
The length of the cell $L_{cell}$ can be adjusted by moving the back of the cell inside the cylinder, similarly to a plunger. The maximum length is $\sim10$ mm and the minimum of the order of 100 $\upmu$m inner length.\\
The gas inlet is connected to a reservoir located outside the experimental chamber and isolated by an electro-valve. The electro-valve opening allows the gas to expand to the cell and flow through the plates apertures into the experimental chamber. The valve opening duration $\tau_{valve}$ can be varied, down to a minimum duration of $\sim 30$ ms.

\section{Experimental characterization}
\label{S:3}

\subsection{Interferometric setup}
\label{S:3.1}
Prior to laser wakefield experiments, the averaged gas density was characterized experimentally with the use of a Mach-Zender interferometer with a setup similar to the one in ref. \cite{Ju2012}. A helium-neon laser with wavelength $\lambda_0 \simeq 632$ nm is filtered and splitted into two beams. A reference beam propagates in a $\sim10^{-5}$ mbar vacuum while a probe beam is focused transversely inside the cell through its windows before being recollimated. The two beams are then recombined and an interference pattern is formed. The light corresponding to about $\sim1/3$ of the brightest fringe is selected by a slit and focused onto a photodiode. The photodiode signal changes with time when gas is let in the cell, and is recorded with an oscilloscope. During a typical measurement sequence, the valve is opened at $t=0$ ms, the gas then expands from the reservoir and reaches the cell after $\sim30$ ms. The phase difference between the probe beam and the reference beam is increased by the gas filling the cell; the phase variation is calculated from the variation of the intensity going through the slit and focused onto the diode. The quantity measured is the phase shift :
\begin{equation}
\Delta \varphi(t) \simeq (2\pi/\lambda_0) \int_L (\eta(z,t)-1) dz
\label{eq:Eq1}
\end{equation} 
with $\eta$ the refractive index of the gas ($\mathrm{H_2}$) which is linked to the gas density and $L$ the transverse length of the cell. 
As we can see from Eq. \ref{eq:Eq1}, this measurement is integrated along the probe beam path inside the cell and does not provide information on the spatial distribution. An homogeneous density distribution along the probe beam path is assumed which is confirmed to be reasonable by fluid simulation results (see Sec.\ref{S:4.1}).

\subsection{Density measurements}
\label{S:3.2}

\begin{figure}[h]
\centering
\includegraphics[scale=0.22]{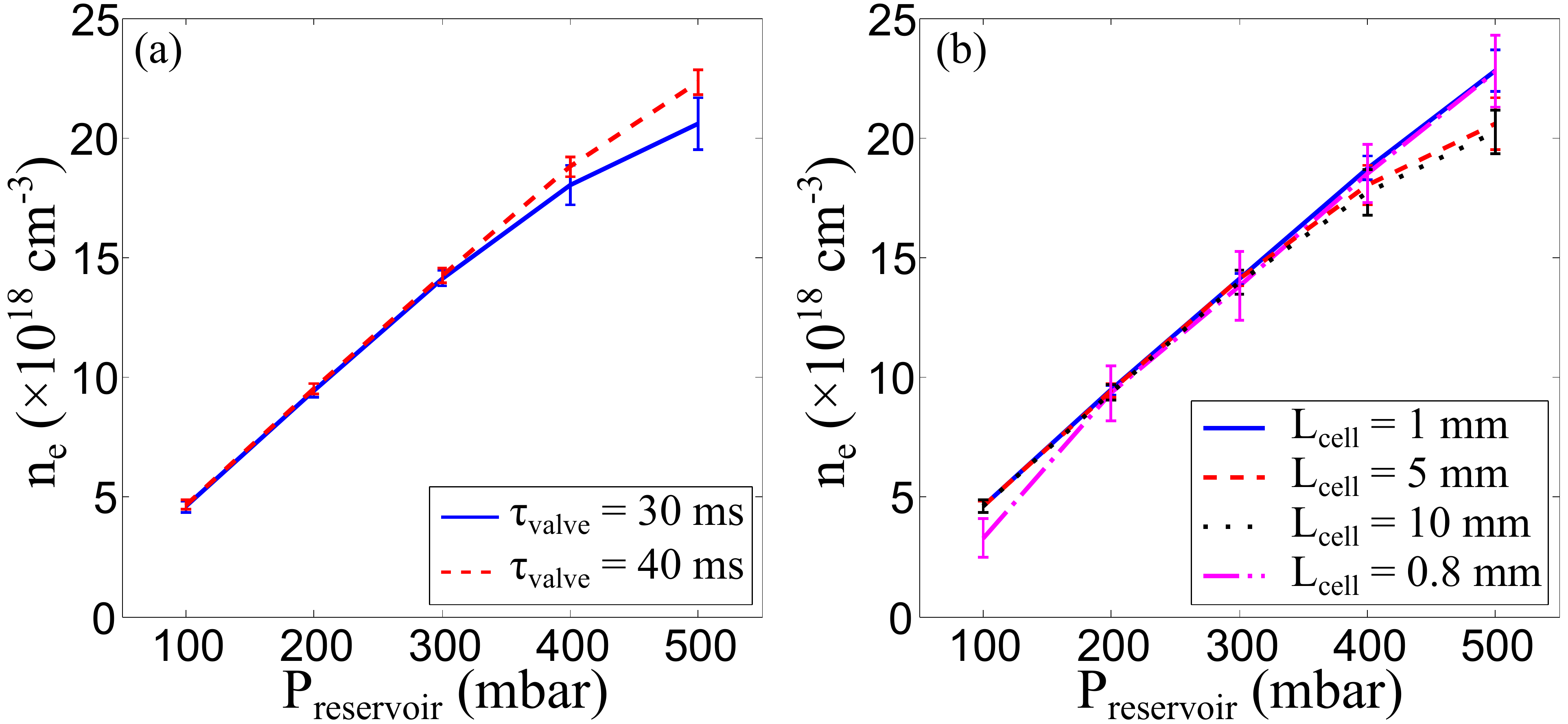}
\caption{(a) Mean density inside the cell as a function of the reservoir backing pressure for $\tau_{valve} = 30$ ms and $\tau_{valve} = 40$ ms with $L_{cell} = 5$ mm.(b) Mean density inside the cell as a function of the reservoir backing pressure for $L_{cell} = 1$, $5$ and $10$ mm with $\tau_{valve} = 40$ ms, $d_{plate}=0.5$ mm and $\phi_{aperture}=0.2$ mm and for $L_{cell} = 0.8$ mm with $\tau_{valve} = 40$ ms, $d_{plate}=0.5$ mm and larger apertures of $\phi_{aperture}=0.6$ mm. \label{fig:Fig2}}
\end{figure}

Results of interferometric measurements are presented in Fig. \ref{fig:Fig2} (a-b). Fig.\ref{fig:Fig2} (a) shows a linear dependence between the mean density inside the cell and the backing pressure, similar for $\tau_{valve}=30$ ms and $\tau_{valve}=40$ ms in the range of electronic density $n_e\sim[4.7 ; 14.3]\times10^{18}$ $\mathrm{cm^{-3}}$. Above this range, the slope is decreased and a valve opening duration of $\tau_{valve}=40$ ms allows for a higher density and lower density fluctuations.
In Fig. \ref{fig:Fig2} (b), the mean density is displayed as a function of the backing pressure for $L_{cell}=1$, $5$ and $10$ mm for $\phi_{aperture}=0.2$ mm. No difference can be seen in the mean density when the cell length is varied for a given backing pressure up to $P_{reservoir}=500$ mbar for which smaller cell length sustain higher density (of the order of $\sim8$ \% higher for $L_{cell} = 1$ mm compared to $L_{cell} = 10$ mm).

Comparison of interferometric measurements for $\phi_{aperture} = 0.2$ mm and $\phi_{aperture}=0.6$ mm is shown in Fig. \ref{fig:Fig2}(b) for the cases $L_{cell}=1$ mm and $L_{cell}=0.8$ mm. It seems to indicate that an aperture of 0.2 mm allows for slightly higher densities inside the cell but error bars overlap. Smaller fluctuations for a given backing pressure are also observed for 0.2 mm apertures. However, 0.2 mm apertures can be ablated by the high intensity laser during laser wakefield experiments. Hence, 0.6 mm diameter apertures should be preferred to ensure stable experimental conditions.

\section{Fluid simulations}
\label{S:4}
Fluid simulations were performed with the OpenFOAM software and using the sonicFoam solver. The sonicFoam solver is a transient, turbulent solver with sonic flow capabilities which is suitable for pressure differences between the reservoir (and at later times the inside of the cell) and the vacuum of the experimental chamber of at least $10^3$ mbar.

\subsection{Full geometry simulation}
\label{S:4.1}
The first approach in simulations was to calculate the gas flow inside a full 3D geometry approximation of the cell as accurately as possible to compare the calculated density value to the experimental one and to test the hypothesis of a flat transverse density profile.

\begin{figure}[h]
\centering
\includegraphics[scale=0.45]{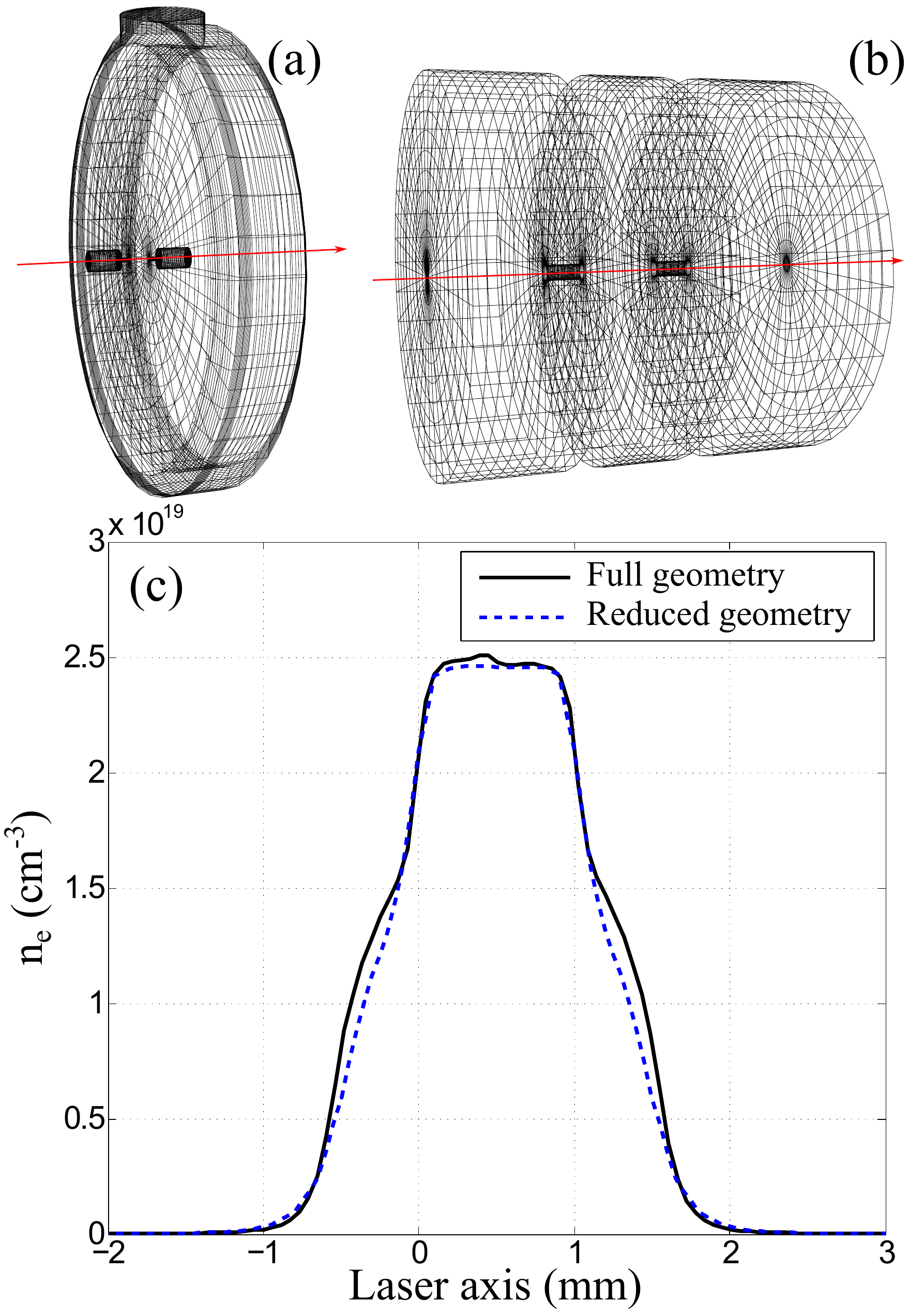}
\caption{(a) Full 3D geometry of the ELISA gas cell. The red arrow represents the high power laser axis for LWFA experiments. (b) Reduced geometry representing the surroundings of the cylinder axis. The red arrow represents the high power laser axis for LWFA experiments. (c) Normalized density profile along the laser for the full geometry and the reduced geometry with the same pressure inlet area. In both cases $L_{cell}=1$ mm, $\phi_{aperture}=0.2$ mm, $d_{plate}=0.5$ mm and $P_{reservoir}=500$ mbar.\label{fig:Fig3}}
\end{figure}

The full geometry is shown in Fig. \ref{fig:Fig3}(a). In this example $L_{cell}=1$ mm, so the inner cell length is actually shorter than the diameter of the gas inlet tube at the top. The gas inlet flows inside a groove just below the gas inlet tube and surrounding the cell. This groove allows the gas to flow around the cylinder as well as directly into its central part. 
The boundary conditions used in this simulation are a fixed pressure on the gas inlet tube surface, transmissive conditions on vacuum box surfaces (to keep their size small without filling them and changing the gas flow) and wall boundary conditions everywhere else.
The pressure inside the cell increases and even exceeds the inlet pressure and then decreases and reaches a stationary value. 
In the stationary state, the difference of the integrals  between a flat density profile and the simulated transverse density profile is of the order of 0.4 \% over the cell diameter (20 mm), corresponding to the probe beam path in the experiment. Thus the hypothesis of a flat density profile in the analysis of the interferometric measurements is reasonable. 
The mean transverse electronic density, assuming complete ionization, for $L_{cell}=1$ mm, $\phi_{aperture}=0.2$ mm, $d_{plate} = 0.5$ mm and $P_{reservoir}=500$ mbar is $<n_e>\simeq 2.53\times 10^{19}$ $\mathrm{cm^{-3}}$. In order to compare to experimental values, gas expansion into a connection tube has to be taken into account, as the reservoir pressure is applied directly at the entry of the cell in simulation whereas in the experiment, the gas undergoes an expansion from the reservoir to the cell. If we take into account the ratio of volume before and after the valve opening, a transverse mean value of $<n_e>^c\simeq 0.95\times <n_e>\simeq 2.4\times 10^{19}$ $\mathrm{cm^{-3}}$ is obtained. This value is in very good agreement with the experimental values of $n_e\simeq (2.4 \pm 0.05)\times 10^{19}$ $\mathrm{cm^{-3}}$ obtained in the same conditions with $\tau_{valve}=40$ ms.

Despite the good agreement between this simulation and the experimental results, the computational time requested by the simulation of the whole geometry is incompatible with a parametric study. 

\subsection{Parametric study in reduced geometry}
\label{S:4.2}
A reduced geometry limited to the surroundings of the cylinder axis was implemented to perform a parametric study. It consists of an ensemble of 3 cylinders of 5 mm diameter representing the vacuum boxes and the inside of the cell connected to each others by smaller diameter cylinders representing the plate apertures as illustrated in Fig. \ref{fig:Fig3}(b). On the side of the central cylinder, representing the inner part of the cell, we apply a boundary condition of fixed pressure representing the gas inlet. Other boundary conditions are kept identical to the full geometry case.
Fig. \ref{fig:Fig3} (c) shows that a good agreement between the full geometry and the reduced geometry when the gas inlet area is the same. The local difference does not exceed 10 \% and the difference between the integrals of the profiles is  $\sim 0.8$ \%.

\subsubsection{Influence of the aperture diameter $\phi_{aperture}$}
\label{S:4.2.1}

\begin{figure}[h]
\centering
\includegraphics[scale=0.4]{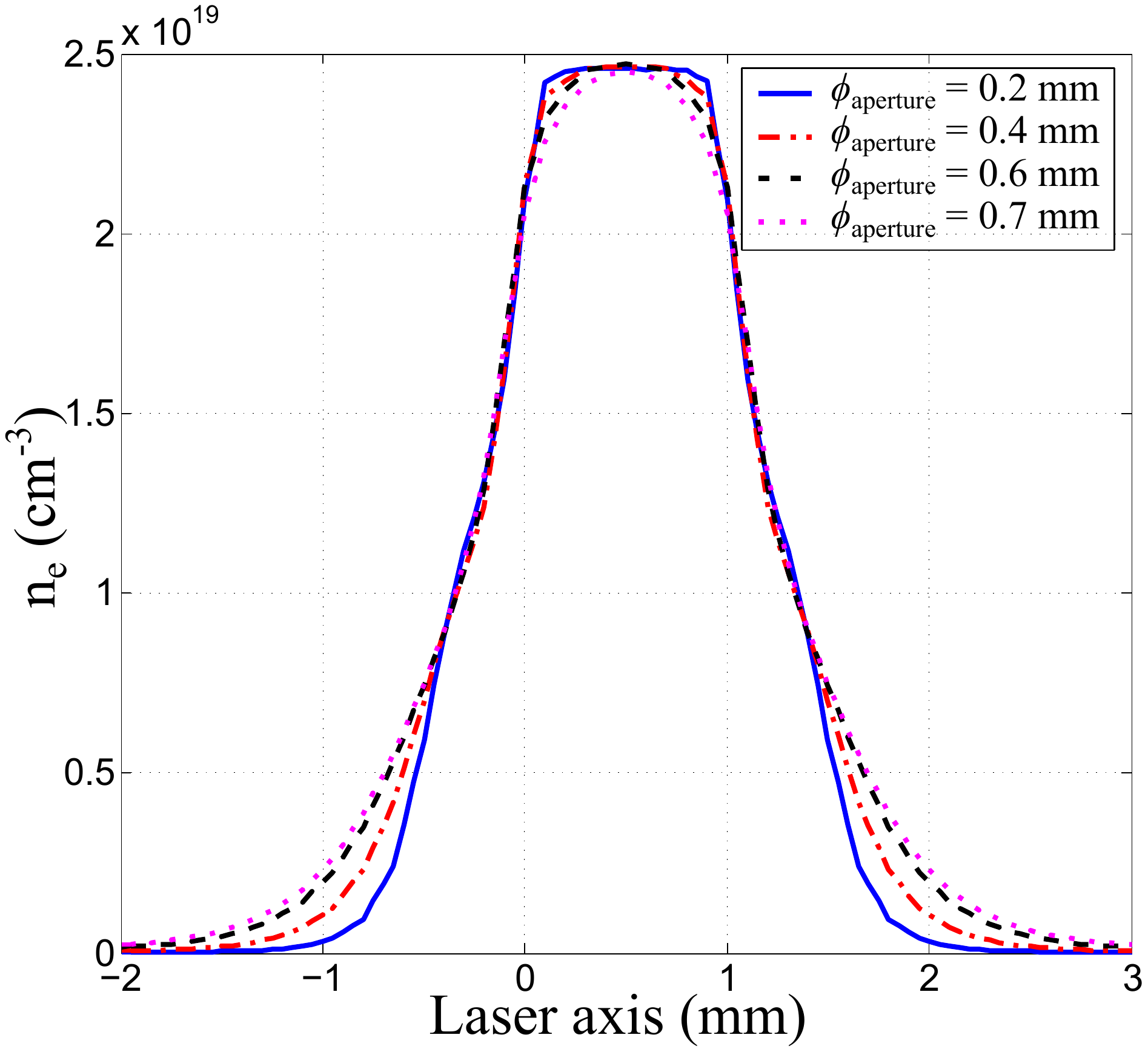}
\caption{Electronic density profile along the laser axis assuming complete ionization for different aperture diameters : $\phi_{aperture}=0.2$, $0.4$, $0.6$ and $0.7$ mm. Other parameters are : $L_{cell} = 1$ mm, $d_{plate}=0.5$ mm and $P_{reservoir}=500$ mbar.\label{fig:Fig4}}
\end{figure}

Changing the diameter of the plate apertures allows one to control the shape of the gradients of the density profile along the laser axis. 
In Fig. \ref{fig:Fig4} we plot the electronic density profiles (assuming complete ionization) along the central axis of the cell for different values of $\phi_{aperture}$. Other parameters are kept constant and are as follows : $L_{cell}=1$ mm, entry and exit plates are identical and their thickness is $d_{plate}=0.5$ mm and $P_{reservoir}=500$ mbar.
The density profiles are shown at the end of each simulation when the stationary state is reached. Density values at the plateau during the stationary states are very similar as the standard deviation is  only $0.15$\% of the mean value.
We can see that when $\phi_{aperture}$ is increased, the gradients are smoother and extend further outside the cell. The plateau inside the cell is also shortened so that the profile with large apertures ($\phi_{aperture}\geq0.6$ mm) is closer to a bell-shaped curve than to a plateau surrounded by gradients as it is the case for smaller apertures ($\phi_{aperture}\leq0.4$ mm).

\subsubsection{Influence of the plate thickness $d_{plate}$}
\label{S:4.2.2}

The other parameter to change the gradient shape is the plate thickness $d_{plate}$. The effect of a thicker plate on the density profile is explored by increasing only the exit plate thickness up to $d_{plate}=2$ mm but keeping the entry plate at $d_{plate}=0.5$ mm.

\begin{figure}[h]
\centering
\includegraphics[scale=0.4]{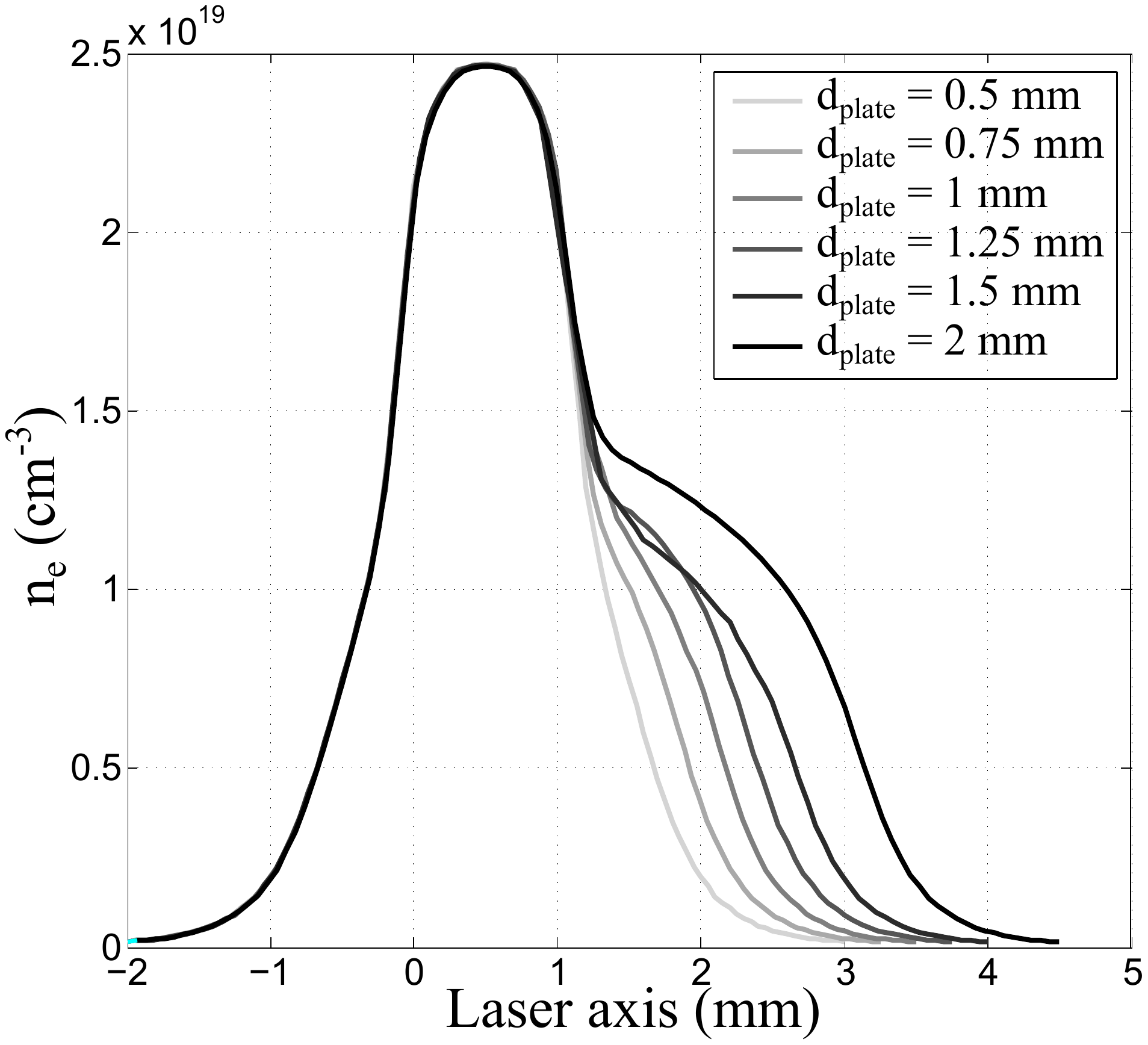}
\caption{Electronic density profile along the laser axis assuming complete ionization for different exit plate thickness : $d_{plate}=0.5$, $0.75$, $1$, $1.25$, $1.5$ and $2$ mm. other parameters are $L_{cell}=1$ mm, $\phi_{aperture}=0.6$ mm. \label{fig:Fig5}}
\end{figure}

Several electronic density profiles are plotted (assuming complete ionization) in Fig. \ref{fig:Fig5} for different exit plate thickness. Other parameters are kept constant and are : $\phi_{aperture}=0.6$ mm, $L_{cell}=1$ mm and $P_{reservoir}=500$ mbar. The profiles plotted are obtained at the stationary state which is reached in each simulations. In all these cases the maximum values of density are very similar (standard deviation of $\sim 0.05$ \%). 
Fig. \ref{fig:Fig5} shows that a thicker exit plate extends the gradients and creates an inflection at the transition between the inner part of the cell and the tube formed by the plate (around the position $1$ mm in Fig. \ref{fig:Fig5}). The density profile then exhibits an "elbow" at around half of the maximum density which can be beneficial for further acceleration in LWFA experiments.

\subsubsection{Influence of the cell length $L_{cell}$}
\label{S:4.2.3}
\begin{figure}[h]
\centering
\includegraphics[scale=0.4]{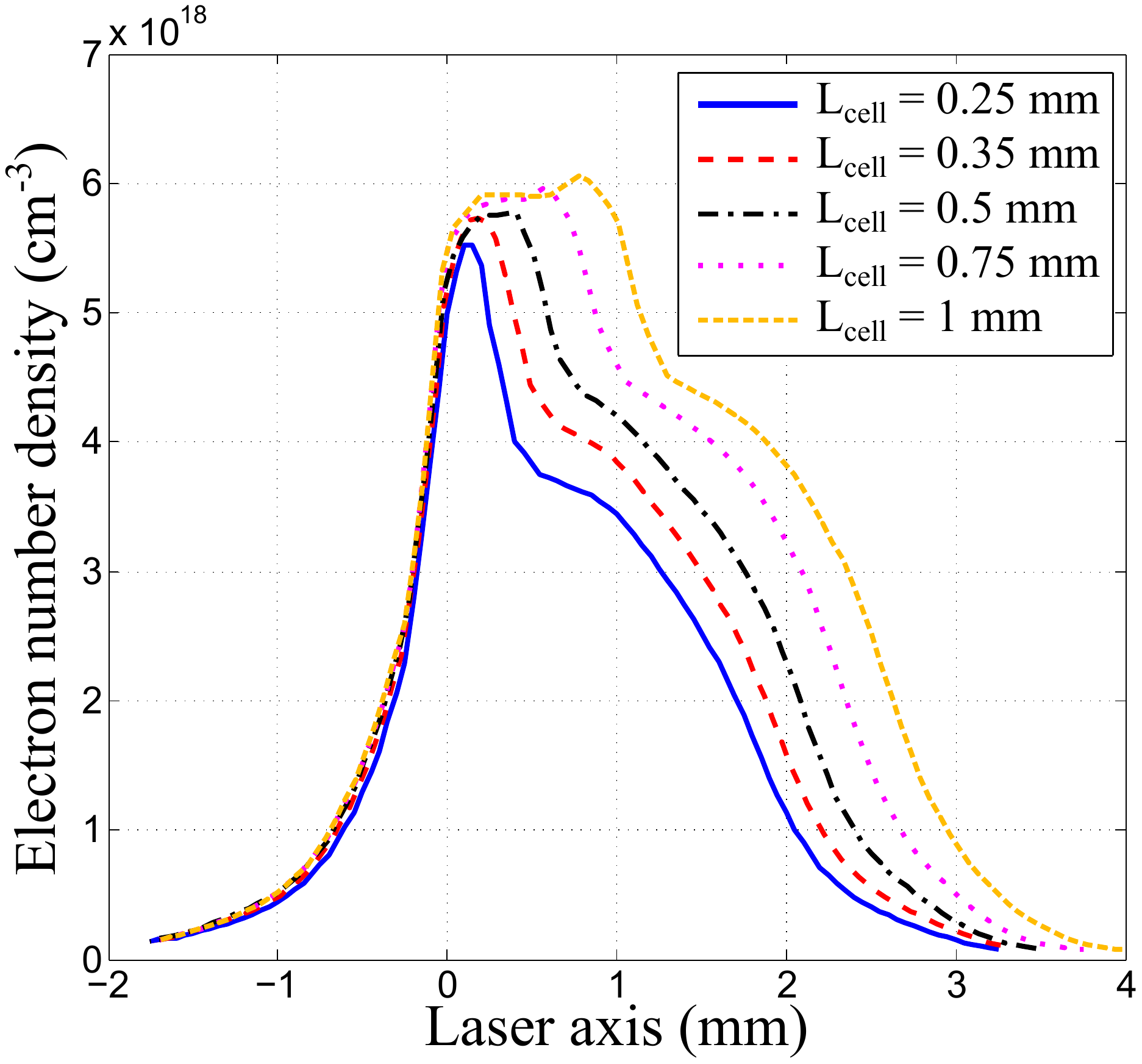}
\caption{Electronic density profile along the laser axis assuming complete ionization for different cell length : $L_{cell} = 0.25$, $0.35$, $0.5$, $0.75$ and $1$ mm. Other parameters are : $\phi_{aperture}=0.6$ mm, entry plate with $d_{plate}=0.25$ mm, exit plate with $d_{plate}=1.5$ mm and $P_{reservoir} = 120$ mbar. \label{fig:Fig6}}
\end{figure}

Fig. \ref{fig:Fig6} presents several electronic density profiles (assuming complete ionization) for different cell lengths and all other parameters fixed. It shows that the cell length determines the density plateau extension but also has an impact both on the maximum density value and the gradient. 
When the cell length is extended, the maximum density is slightly increased (from $5.52\times10^{18}$ $\mathrm{cm^{-3}}$ for $L_{cell}=0.25$ mm to $6.06\times10^{18}$ $\mathrm{cm^{-3}}$ for $L_{cell}=1$ mm). In this case, $P_{reservoir}$ should be adjusted when the cell length is changed to retrieve the same maximum density. 
The value of $n_e$ where the exit gradient changes is also slightly modified, even compared to the relative maximum : it can be observed at $\sim0.68$ of the maximum density for $L_{cell}=0.25$ mm and at $\sim0.74$ of the maximum density for $L_{cell}=0.25$ mm. This could change the evolution of a high power laser interacting with a plasma with such a density distribution. 

\section{Conclusion}

A variable parameter gas cell was designed and characterized both experimentally, by the means of interferometry, and by fluid simulations. Experimental results show a linear dependence between the mean density inside the cell and the backing pressure for all values of cell length. Fluid simulations were performed to investigate the density distribution inside the cell. The mean density obtained in simulations is in good agreement with experimental results an the geometry provides achievable density profiles along the laser axis relevant to LWFA experiments. 
A parametric study was performed by fluid simulations in a reduced geometry. It provides simple and robust solutions to change the gradient length, the plateau extension and the possibility to add a second plateau at around half of the maximum density inside the cell.
These density distributions will be used as input for particle-in-cell simulations to find an optimum working point.

\subsection*{Acknowledgement}
This work was supported by the Triangle de la Physique under contract no 2012-032TELISA. T. L. Audet acknowlegdes financial support of EuPRAXIA, co-funded by the European Commission in its Horizon2020 Programme under the Grant Agreement no 653782.

\bibliographystyle{model1-num-names}
\bibliography{bibFile}

\begin{thebibliography}{11}
\expandafter\ifx\csname natexlab\endcsname\relax\def\natexlab#1{#1}\fi
\providecommand{\bibinfo}[2]{#2}
\ifx\xfnm\relax \def\xfnm[#1]{\unskip,\space#1}\fi
\bibitem[{Spence and Hooker(2000)}]{Spence2000}
\bibinfo{author}{D.~J. Spence}, \bibinfo{author}{S.~M. Hooker},
\newblock \bibinfo{title}{Investigation of a hydrogen plasma waveguide},
\newblock \bibinfo{journal}{Phys. Rev. E} \bibinfo{volume}{63}
  (\bibinfo{year}{2000}) \bibinfo{pages}{015401}.
\bibitem[{Ammosov et~al.(1986)Ammosov, Delone, and Krainov}]{Ammosov1986}
\bibinfo{author}{M.~V. Ammosov}, \bibinfo{author}{N.~B. Delone},
  \bibinfo{author}{V.~P. Krainov},
\newblock \bibinfo{title}{Tunnel ionization of complex atoms and of atomic ions
  in an alternating electromagnetic field},
\newblock \bibinfo{journal}{Sov. Phys. JETP} \bibinfo{volume}{64}
  (\bibinfo{year}{1986}) \bibinfo{pages}{1191--1194}.
\bibitem[{Semushin and Malka(2001)}]{Semushin2001}
\bibinfo{author}{S.~Semushin}, \bibinfo{author}{V.~Malka},
\newblock \bibinfo{title}{{High density gas jet nozzle design for laser target
  production}},
\newblock \bibinfo{journal}{Review of Scientific Instruments}
  \bibinfo{volume}{72} (\bibinfo{year}{2001}) \bibinfo{pages}{2961}.
\bibitem[{Andreev et~al.(2010)Andreev, Cassou, Wojda, Genoud, Burza, Lundh,
  Persson, Cros, Fortov, and Wahlstrom}]{Andreev2010}
\bibinfo{author}{N.~E. Andreev}, \bibinfo{author}{K.~Cassou},
  \bibinfo{author}{F.~Wojda}, \bibinfo{author}{G.~Genoud},
  \bibinfo{author}{M.~Burza}, \bibinfo{author}{O.~Lundh},
  \bibinfo{author}{A.~Persson}, \bibinfo{author}{B.~Cros},
  \bibinfo{author}{V.~E. Fortov}, \bibinfo{author}{C.-G. Wahlstrom},
\newblock \bibinfo{title}{Analysis of laser wakefield dynamics in capillary
  tubes},
\newblock \bibinfo{journal}{New Journal of Physics} \bibinfo{volume}{12}
  (\bibinfo{year}{2010}) \bibinfo{pages}{045024}.
\bibitem[{Desforges et~al.(2014)Desforges, Paradkar, Hansson, Ju, Senje, Audet,
  Persson, Dufr\'enoy, Lundh, Maynard, Monot, Vay, Wahlstr\"om, and
  Cros}]{Desforges2014}
\bibinfo{author}{F.~G. Desforges}, \bibinfo{author}{B.~S. Paradkar},
  \bibinfo{author}{M.~Hansson}, \bibinfo{author}{J.~Ju},
  \bibinfo{author}{L.~Senje}, \bibinfo{author}{T.~L. Audet},
  \bibinfo{author}{A.~Persson}, \bibinfo{author}{S.~D. Dufr\'enoy},
  \bibinfo{author}{O.~Lundh}, \bibinfo{author}{G.~Maynard},
  \bibinfo{author}{P.~Monot}, \bibinfo{author}{J.-L. Vay},
  \bibinfo{author}{C.-G. Wahlstr\"om}, \bibinfo{author}{B.~Cros},
\newblock \bibinfo{title}{Dynamics of ionization-induced electron injection in
  the high density regime of laser wakefield acceleration},
\newblock \bibinfo{journal}{Phys. Plasmas} \bibinfo{volume}{21}
  (\bibinfo{year}{2014}) \bibinfo{pages}{120703}.
\bibitem[{Clayton et~al.(2010)Clayton, Ralph, Albert, Fonseca, Glenzer, Joshi,
  Lu, Marsh, Martins, Mori, Pak, Tsung, Pollock, Ross, Silva, and
  Froula}]{Clayton2010}
\bibinfo{author}{C.~E. Clayton}, \bibinfo{author}{J.~E. Ralph},
  \bibinfo{author}{F.~Albert}, \bibinfo{author}{R.~A. Fonseca},
  \bibinfo{author}{S.~H. Glenzer}, \bibinfo{author}{C.~Joshi},
  \bibinfo{author}{W.~Lu}, \bibinfo{author}{K.~A. Marsh},
  \bibinfo{author}{S.~F. Martins}, \bibinfo{author}{W.~B. Mori},
  \bibinfo{author}{A.~Pak}, \bibinfo{author}{F.~S. Tsung},
  \bibinfo{author}{B.~B. Pollock}, \bibinfo{author}{J.~S. Ross},
  \bibinfo{author}{L.~O. Silva}, \bibinfo{author}{D.~H. Froula},
\newblock \bibinfo{title}{Self-guided laser wakefield acceleration beyond 1
  {GeV} using ionization-induced injection},
\newblock \bibinfo{journal}{Phys. Rev. Lett.} \bibinfo{volume}{105}
  (\bibinfo{year}{2010}) \bibinfo{pages}{105003}.
\bibitem[{Audet et~al.(2016)Audet, Hansson, Lee, Desforges, Maynard,
  Dobosz~Dufr{\'e}noy, Lehe, Vay, Aurand, Persson, Gallardo~Gonz{\`a}lez,
  Maitrallain, Monot, Wahlstr{\"o}m, Lundh, and Cros}]{Audet2016}
\bibinfo{author}{T.~L. Audet}, \bibinfo{author}{M.~Hansson},
  \bibinfo{author}{P.~Lee}, \bibinfo{author}{F.~G. Desforges},
  \bibinfo{author}{G.~Maynard}, \bibinfo{author}{S.~Dobosz~Dufr{\'e}noy},
  \bibinfo{author}{R.~Lehe}, \bibinfo{author}{J.-L. Vay},
  \bibinfo{author}{B.~Aurand}, \bibinfo{author}{A.~Persson},
  \bibinfo{author}{I.~Gallardo~Gonz{\`a}lez}, \bibinfo{author}{A.~Maitrallain},
  \bibinfo{author}{P.~Monot}, \bibinfo{author}{C.-G. Wahlstr{\"o}m},
  \bibinfo{author}{O.~Lundh}, \bibinfo{author}{B.~Cros},
\newblock \bibinfo{title}{Investigation of ionization-induced electron
  injection in a wakefield driven by laser inside a gas cell},
\newblock \bibinfo{journal}{Physics of Plasmas} \bibinfo{volume}{23}
  (\bibinfo{year}{2016}).
\bibitem[{Walker and et~al.(2017)}]{Walker2017}
\bibinfo{author}{P.~A. Walker}, \bibinfo{author}{et~al.},
\newblock \bibinfo{title}{Horizon 2020 eupraxia design study},
\newblock \bibinfo{journal}{Journal of Physics: Conference Series}
  \bibinfo{volume}{874} (\bibinfo{year}{2017}) \bibinfo{pages}{012029}.
\bibitem[{Cros et~al.(2013)Cros, Paradkar, Davoine, Chanc\'e, Desforges,
  Dufr\'enoy, Delerue, Ju, Audet, Maynard, Lobet, Gremillet, Mora, Schwindling,
  Delferri\`ere, Bruni, Rimbault, Vinatier, Piazza, Grech, Riconda, Marqu\`es,
  Beck, Specka, Martin, Monot, Normand, Mathieu, Audebert, and
  Amiranoff}]{Cros2013}
\bibinfo{author}{B.~Cros}, \bibinfo{author}{B.~S. Paradkar},
  \bibinfo{author}{X.~Davoine}, \bibinfo{author}{A.~Chanc\'e},
  \bibinfo{author}{F.~G. Desforges}, \bibinfo{author}{S.~D. Dufr\'enoy},
  \bibinfo{author}{N.~Delerue}, \bibinfo{author}{J.~Ju}, \bibinfo{author}{T.~L.
  Audet}, \bibinfo{author}{G.~Maynard}, \bibinfo{author}{M.~Lobet},
  \bibinfo{author}{L.~Gremillet}, \bibinfo{author}{P.~Mora},
  \bibinfo{author}{J.~Schwindling}, \bibinfo{author}{O.~Delferri\`ere},
  \bibinfo{author}{C.~Bruni}, \bibinfo{author}{C.~Rimbault},
  \bibinfo{author}{T.~Vinatier}, \bibinfo{author}{A.~D. Piazza},
  \bibinfo{author}{M.~Grech}, \bibinfo{author}{C.~Riconda},
  \bibinfo{author}{J.~R. Marqu\`es}, \bibinfo{author}{A.~Beck},
  \bibinfo{author}{A.~E. Specka}, \bibinfo{author}{P.~Martin},
  \bibinfo{author}{P.~Monot}, \bibinfo{author}{D.~Normand},
  \bibinfo{author}{F.~Mathieu}, \bibinfo{author}{P.~Audebert},
  \bibinfo{author}{F.~Amiranoff},
\newblock \bibinfo{title}{Laser plasma acceleration of electrons with
  multi-{PW} laser beams in the frame of {CILEX}},
\newblock \bibinfo{journal}{Nucl. Instrum. Meth. A} \bibinfo{volume}{740}
  (\bibinfo{year}{2013}) \bibinfo{pages}{158 --164}.
\bibitem[{Lee et~al.(2017)Lee, Maynard, Audet, Lehe, Vay, and Cros}]{Lee2017}
\bibinfo{author}{P.~Lee}, \bibinfo{author}{G.~Maynard},
  \bibinfo{author}{T.~Audet}, \bibinfo{author}{R.~Lehe},
  \bibinfo{author}{J.~Vay}, \bibinfo{author}{B.~Cros},
\newblock \bibinfo{title}{Optimization of laser-plasma injector via beam
  loading effects using ionization-induced injection},
\newblock \bibinfo{journal}{arXiv preprint arXiv:1711.01613}
  (\bibinfo{year}{2017}).
\bibitem[{Ju and Cros(2012)}]{Ju2012}
\bibinfo{author}{J.~Ju}, \bibinfo{author}{B.~Cros},
\newblock \bibinfo{title}{Characterization of temporal and spatial distribution
  of hydrogen gas density in capillary tubes for laser-plasma experiments},
\newblock \bibinfo{journal}{J. Appl. Phys.} \bibinfo{volume}{112}
  (\bibinfo{year}{2012}) \bibinfo{pages}{113102}.

\end{thebibliography}

\end{document}